\documentclass{article}

\usepackage{arxiv}

\usepackage[utf8]{inputenc} 
\usepackage[T1]{fontenc}    
\usepackage{hyperref}       
\usepackage{url}            
\usepackage{booktabs}       
\usepackage{amsfonts}       
\usepackage{amsmath}
\usepackage{nicefrac}       
\usepackage{microtype}      
\usepackage{cleveref}       
\usepackage{lipsum}         
\usepackage{graphicx}
\usepackage{natbib}
\usepackage{doi}
\usepackage{titlesec}
\usepackage{tabu}
\usepackage{longtable}
\usepackage{enumitem}
\usepackage{amssymb}
\usepackage{xcolor}
\newlist{selectlist}{itemize}{2}
\setlist[selectlist]{label=$\square$,leftmargin=*,noitemsep,topsep=0pt}
\usepackage{epsfig} 
\usepackage{subfig}
\usepackage{multirow}

\usepackage{lmodern}

\usepackage{hyperref}
\hypersetup{
    colorlinks=true,
    linkcolor=blue,
    filecolor=magenta,      
    urlcolor=blue,
}

\titleformat{\section}[block]{\hspace{1em}\bfseries}{\thesection.}{0.5em}{} 
\titleformat{\subsection}[block]{\hspace{1em}}{\thesubsection}{0.5em}{}

\title{Non-central panorama indoor dataset}

\date{}

\author{ \href{https://orcid.org/0000-0003-2674-4844}{\includegraphics[scale=0.06]{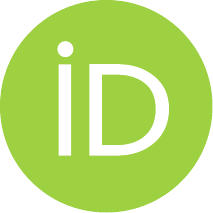}\hspace{1mm}Bruno Berenguel-Baeta}\thanks{Corresponding author.} \\
	Instituto de Investigacion en Ingenieria de Aragon\\
	Department of Computer Science and Systems Engineering\\
	University of Zaragoza,
	Zaragoza, Spain \\
	\texttt{berenguel@unizar.es} \\
	\And
	\href{https://orcid.org/0000-0002-8479-1748}{\includegraphics[scale=0.06]{orcid.pdf}\hspace{1mm}Jesus Bermudez-Cameo} \\
	Instituto de Investigacion en Ingenieria de Aragon\\
	Department of Computer Science and Systems Engineering\\
	University of Zaragoza,
	Zaragoza, Spain \\
	\texttt{bermudez@unizar.es} \\
	\And
	\href{https://orcid.org/0000-0001-5209-2267}{\includegraphics[scale=0.06]{orcid.pdf}\hspace{1mm}Jose J. Guerrero} \\
	Instituto de Investigacion en Ingenieria de Aragon\\
	Department of Computer Science and Systems Engineering\\
	University of Zaragoza,
	Zaragoza, Spain \\
	\texttt{josechu.guerrero@unizar.es} \\
}
\renewcommand{\shorttitle}{\textit{Floor extraction and door detection}}
\newcommand\blfootnote[1]{%
  \begingroup
  \renewcommand\thefootnote{}\footnote{#1}%
  \addtocounter{footnote}{-1}%
  \endgroup
}
\hypersetup{
pdftitle={Non-central panorama indoor dataset},
pdfsubject={cs.CV},
pdfauthor={David S.~Hippocampus, Elias D.~Striatum},
pdfkeywords={First keyword, Second keyword, More},
}

\begin{document}
\maketitle
\blfootnote{A final version of this article can be found at \url{https://doi.org/10.1016/j.dib.2022.108375}}

\begin{abstract}

Omnidirectional images are one of the main sources of information for learning based scene understanding algorithms. However, annotated datasets of omnidirectional images cannot keep the pace of these learning based algorithms development. 
Among the different panoramas and in contrast to standard central ones, non-central panoramas provide geometrical information in the distortion of the image from which we can retrieve 3D information of the environment [2]. However, due to the lack of commercial non-central devices, up until now there was no dataset of these kinds of panoramas. In this data paper, we present the first dataset of non-central panoramas for indoor scene understanding.
The dataset is composed by {\bf 2574} RGB non-central panoramas taken in around 650 different rooms. Each panorama has associated a depth map and annotations to obtain the layout of the room from the image as a structural edge map, list of corners in the image, the 3D corners of the room and the camera pose. The images are taken from photorealistic virtual environments and pixel-wise automatically annotated.
\end{abstract}
\newpage
\begin{flushleft}

\textbf{Specifications Table}\\
\vskip 0.2cm 
\begin{longtable}{|p{33mm}|p{124mm}|}
\hline
\textbf{Subject}                & Computer Science: Computer Vision and Pattern Recognition\\
\hline                         
\textbf{Specific subject area}  & Non-central circular panoramas for indoor scene understanding. \\
\hline
\textbf{Type of data}           &   RGB Image (.png) \newline
 									Color code depth maps (.png) \newline 
 									Layout annotations (.png, .npy, .txt, .mat)
 									\\                                   
\hline
\textbf{How the data were acquired} & Random generation of virtual environments. \newline
									  Ad-hoc programmable camera projection model for image rendering via ray tracing. The RGB images are rendered with POV-Ray\footnote{The Persistence of Vision Raytracer.\url{http://www.povray.org} (accessed May 2022)} and the depth maps with Mega-POV\footnote{MegaPOV. \url{http://megapov.inertart.net} (accessed May 2022)}. \newline
									  Layout annotations are obtained from the 3D model of the virtual environment.\\
\hline                         
\textbf{Data format}            &   Raw \newline
									Filtered
\\                                                    
\hline
\textbf{Description of          
data collection}             &  From the generated virtual environments, we randomly place the non-central acquisition system in different locations inside the environment. The radius of acquisition is 1 meter. The size of the panoramas is 1024x512 pixels. Once acquired the non-central panoramas, we exclude those that will be physically imposible to acquire in a real situation (i.e. the non-central acquisition system goes through an object or a wall, creating a black hole in the image). 
\\                         
\hline                         
\textbf{Data source location}   & Institution: University of Zaragoza, Department of Computer Science and Systems Engineering \newline
City/Town/Region: Zaragoza, Aragon\newline
Country: Spain\newline
Latitude and longitude for collected samples/data: 41.68390649378424 N -0.8887470805962938 E\\
\hline                         
\hypertarget{target1}
{\textbf{Data accessibility}}   & Repository name: Google Drive (via Github repository to make it easier for the user) \newline
URL to Github repository: \url{https://github.com/jesusbermudezcameo/NonCentralIndoorDataset} \newline
URL to data (Google Drive shared folder): \url{https://drive.google.com/drive/folders/18OQXpbZsr3RBphU0kJC0OS2OXr-3BrkV?usp=sharing}
\\                         
\hline                         
\textbf{Related research\newline
		article}                & B. Berenguel-Baeta, J. Bermudez-Cameo and J.J. Guerrero, Atlanta Scaled Layouts from Non-central Panoramas. Pattern Recognition (2022). DOI:\url{https://doi.org/10.1016/j.patcog.2022.108740}
\\
\hline                         
\end{longtable}


\textbf{Value of the Data}\\
\begin{itemize}
\itemsep=0pt
\parsep=0pt
\item[$\bullet$]The presented dataset is the first existing dataset with non-central panoramas. Besides it includes annotations for different purposes as layout recovery, line extraction and depth estimation.
\item[$\bullet$]Researchers who want to take advantage of the geometrical properties of non-central systems can find in this dataset a perfect source of information for evaluation and development of new algorithms.
\item[$\bullet$]Since it is the first non-central dataset, it can be used to adapt existing algorithms for omnidirectional central images to the no-central case. Besides, in the related research [1], only the RGB images and layout annotations have been used, leaving the depth maps for future research topics.
\end{itemize}
\vskip 0.5cm

\textbf{Data Description}\\
The dataset contains a set of gravity oriented panoramas. We make this specification since non-central panoramas cannot be rotated as a data augmentation in a different axis that the revolution axis of the non-central system. The dataset includes the folders: \textit{img} contains the RGB non-central panoramas are located; \textit{depth\_coded} contains depth maps coded in 3 channels (RGB channels) associated with the RGB panoramas; \textit{EM\_gt} contains one channel images where the structural lines of the environments are defined; \textit{DataPython} and \textit{DataMatLab} contains ground truth information used to evaluate the work presented in [1], including the 3D position of the corners of the room (\textsl{3D\_gt}), the camera location (\textsl{cam\_pose}), the labelling for the floor-wall and ceiling-wall intersections as spherical coordinates of the projecting rays (\textsl{label\_ang}) and the pixel coordinates of the 3D corners in the panoramas (\textsl{label\_cor}).

The main folder follows the following distribution. 
\begin{itemize}
	\item \texttt{NonCentralIndoorDataset}
	\begin{itemize}
		\item \textup{img}
		\item \textup{depth\_coded}
		\item \textup{EM\_gt}
		\item \textup{DataPython}
		\begin{itemize}
			\item \textsl{3D\_gt}
			\item \textsl{cam\_pose}
			\item \textsl{label\_ang}
			\item \textsl{label\_cor}
		\end{itemize}
		\item \textup{DataMatLab}
		\begin{itemize}
			\item \textsl{mat\_gt}
		\end{itemize}
	\end{itemize}		
\end{itemize}

\vskip 0.5cm

\textbf{Experimental design, materials and methods}\\

The data of this dataset is obtained from environments randomly and synthetically generated. We first generate a random layout constrained by different structural limits (see Table \ref{tab:setUpLimits}). These limits include minimum and maximum: area of the room, number of walls, wall length, angle between walls (if non-Manhattan), room height and Manhattan ratio.

\begin{table}[h]
    \caption{Layout limits for randomized generation.}
    \centering
    \begin{tabular}{lcc}
         Parameter 			&  Min. & Max. 	\\ \hline \noalign{\smallskip}
         Area $[m^2]$ 		& 25 	& 110 	\\
         n. walls 			& 4 	& 14 	\\
         wall length $[m]$	& 0.5 	& 8		\\
         walls angle $[degs]$ & 25 	& 100	\\ 
         Room height $[m]$ 	& 2.5 	& 4.25 	\\
         Manhattan ratio 	& \multicolumn{2}{c}{0.7} \\ \hline 
    \end{tabular}
    \label{tab:setUpLimits}
\end{table}

The first version of each layout is a Manhattan layout constrained by the previous limits. Then, depending on the Manhattan ratio, a set of random vertical planes clip the layout introducing oblique walls in angles with respect to the previous walls in the range of the walls angle limit. After the layout clip, we evaluate the final layout, checking if it meets the constraints. If any of the constraints is not satisfied, the layout is deleted and a new one is generated.

Once the structure of the room is obtained, we randomly set different kinds of walls containing doors, windows, colors and textures. These characteristics of the walls are randomly selected from different pools (e.g. we have different models of doors and windows to select). At this point we select the color and textures of the ceiling and floor of the room.
Once defined the 3D structure of the room, we include objects in it. For that purpose, we build a free space map where we can place different kinds of objects. First we consider the kind of objects that are placed next to a wall in a fixed orientation (beds, wardrobes, desks). Then we place objects in the room in a random position and orientation (chairs, sofas, carpets). Finally, we place objects that are placed on top of other objects (cups, clocks, clothes) and we also place lighting sources for a more realistic rendering. All of these objects are randomly picked from different pools depending on the object class. Besides, the ambient illumination conditions are also randomly picked from 3 different configurations.

\captionsetup[subfigure]{labelformat=empty, labelsep=none}
\begin{figure}[t]
	\centering
    \subfloat{\includegraphics[width=0.18\linewidth]{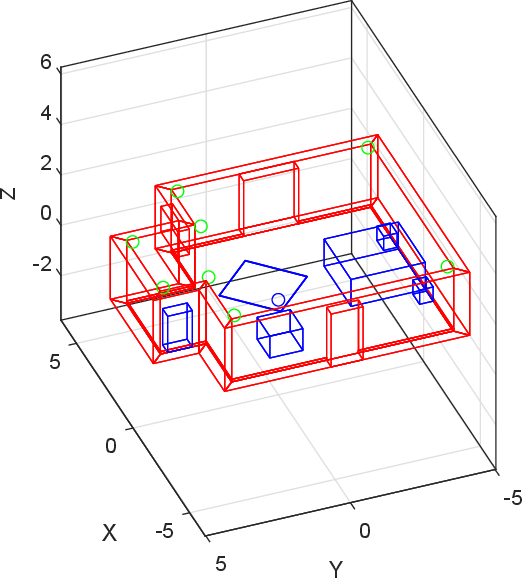}}
	\hfil
    \subfloat[(a)]{\includegraphics[width=0.21\linewidth]{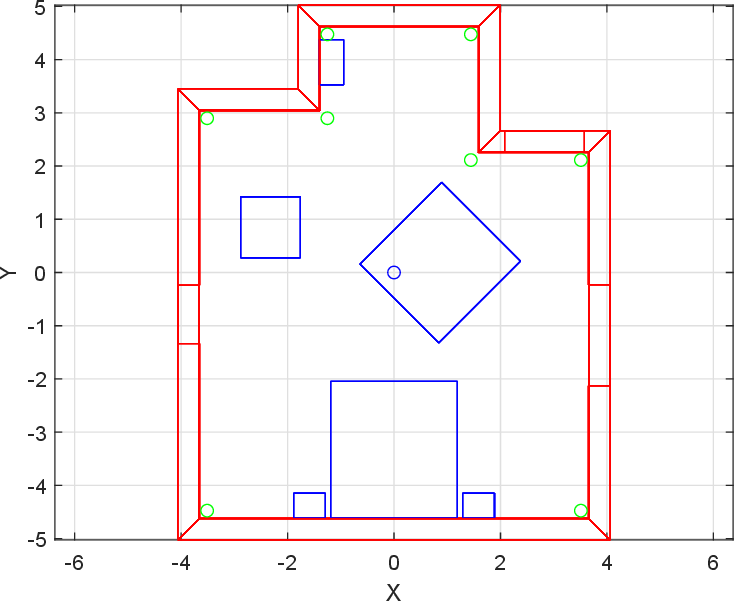}}
    \hfil
    \subfloat{\includegraphics[width=0.40\linewidth]{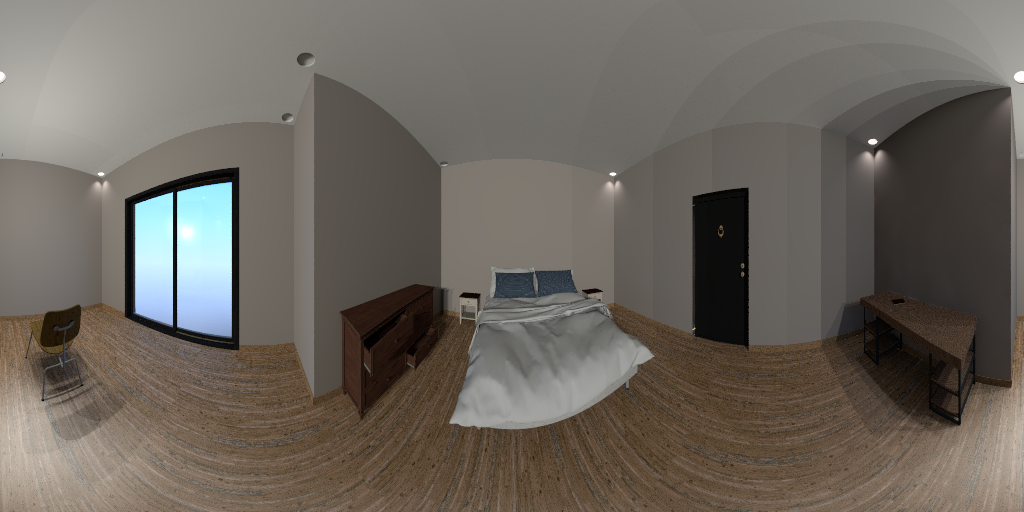}} 
    \\ \vspace{-0.75cm}
    \subfloat{\includegraphics[width=0.18\linewidth]{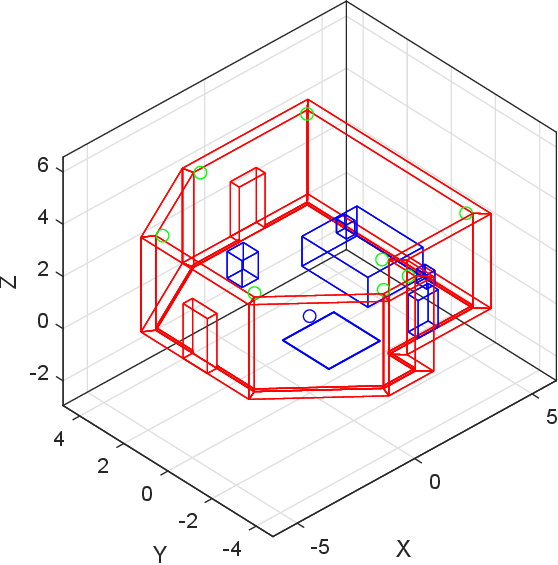}}
    \hfil
    \subfloat[(b)]{\includegraphics[width=0.21\linewidth]{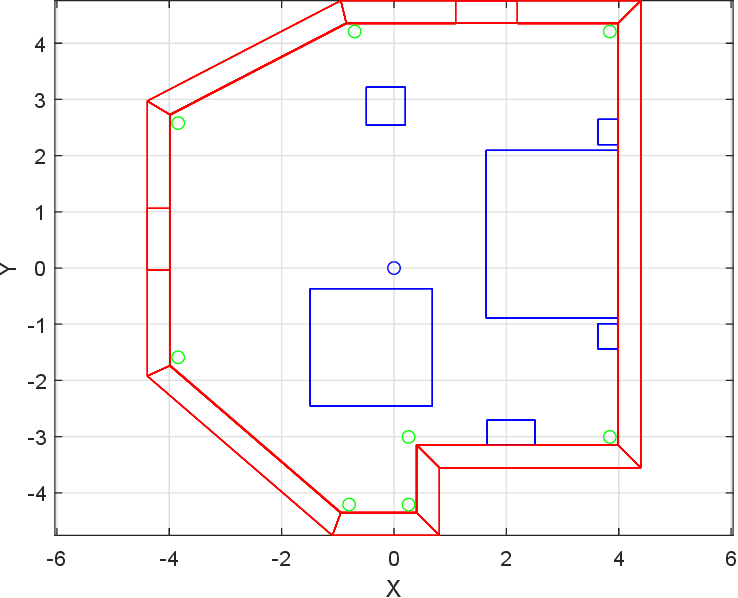}}
    \hfil
    \subfloat{\includegraphics[width=0.40\linewidth]{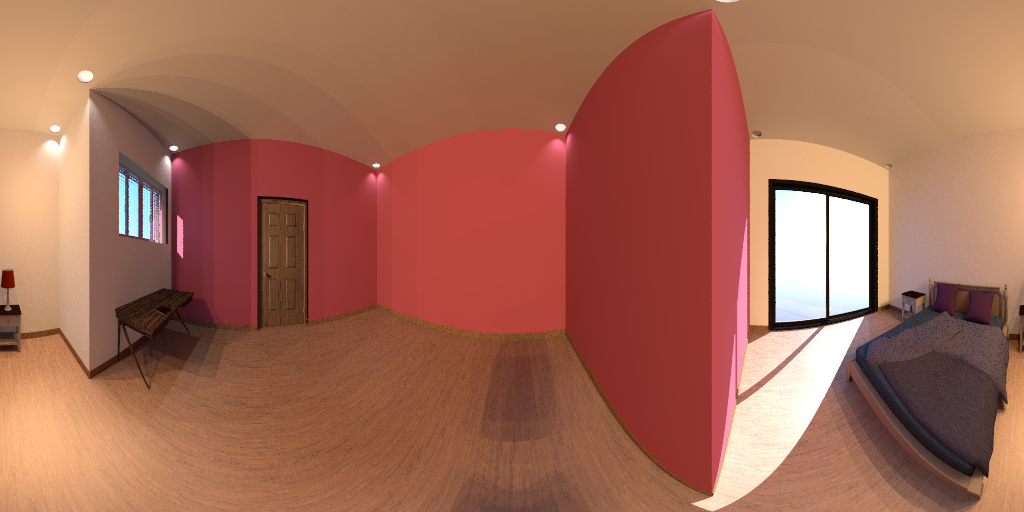}} 
    \caption{Layout generation and non-central panorama rendering examples. (a) Manhattan random room. (b) Atlanta random room.}
    \label{fig:layoutGenerationExamples}
\end{figure}

Once defined our room, we have all the necesary information to generate the ground truth and the labelling of the dataset (see Fig. \ref{fig:layoutGenerationExamples}). The next step is to render the RGB non-central panoramas and generate the depth maps. The color images are rendered with the ray tracing software POV-Ray while the depth maps are obtained with the use of MegaPOV. The non-central panoramic camera is modeled by using an ad-hoc programmable camera projection model included in last versions of POV-Ray. For each scene we can render different acquisitions modifying the position and the orientation of the camera. In the case of the proposed dataset, the camera is always oriented with the gravity direction. Notice that, by contrast with central panoramas (e.g. equirectangular images), we can not post-process a single render for obtaining different panoramas with different orientations.

\vskip 0.5cm

\textbf{Acknowledgments}\\
Funding: This work was supported by RTI2018-096903-B-100 (AEI/ FEDER, UE).\\
\vskip0.3cm

\textbf{References}\\
\begin{itemize}
\item[${[1]}$] B.Berenguel-Baeta, J.Bermudez-Cameo and J.J.Guerrero. Atlanta scaled layouts from non-central panoramas. Pattern Recognition. (2022). \url{https://doi.org/10.1016/j.patcog.2022.108740}
\item[${[2]}$] J. Bermudez-Cameo, O. Saurer, G. Lopez-Nicolas, J.J. Guerrero and M. Pollefeys. Exploiting line metric reconstruction from non-central circular panoramas. Pattern Recognition Letters, 94, 30-37. (2017). \url{https://doi.org/10.1016/j.patrec.2017.05.006}
\end{itemize}

\end{flushleft}
\end{document}